# Observation of near room temperature thin film superconductivity of atmospherically stable Ag-Au mesoscopic thin film

Sobhan Hazra[1], Sandip Chatterjee[2] and Bhola Nath Pal[1,*]

[1]School of Materials Science and Technology, Indian Institute of Technology (Banaras Hindu University), Varanasi-221005, UP, India,

[2]Department of Physics, Indian Institute of Technology (Banaras Hindu University), Varanasi-221005, UP, India,

*Corresponding Author's e-mail: bnpal.mst@iitbhu.ac.in

**Abstract:** An environmentally stable mesoscopic thin film of Au of certain thickness has been deposited thermally on top of a $Ag^+$ implanted oxide substrate to develop a close to room temperature superconductor. This thin film has been deposited in two different stages. Initially, a sol-gel derived ion conducting metal oxide (ICMO) thin film has been deposited by spin coating. Afterward, $Ag^+$ has been introduced inside ICMO thin film by a chemical method. Following this, a thin layer of Au has been deposited on top of that Ag ion-implanted oxide via thermal evaporation. The temperature dependent resistivity (R-T) has been studied by four probe method. During high-to-low temperature sweep, around 240 K this thin film sample shows a sudden drop of resistance from 0.7 Ω to 0.1 μΩ. This 6-7 orders drop of resistance has been observed instantly within <0.1 K temperature variation of the sample. This transition temperature ($T_C$) has been shifted toward the higher temperature by 5-6 degrees when temperature has been increased from low to the higher side. During 2$^{nd}$ and 3$^{rd}$ temperature cycling, both these transitions have been shifted by ~10 K towards room temperature w.r.t the earlier. However, after three successive temperature cycles, $T_C$ becomes stable and transitions occur close to 0 °C repeatedly. At the low resistance phase, current level has been varied from +100 mA to -100 mA which shows a random fluctuation of voltage drop within ±10 nV range, indicating resistance under such circumstance is too low to measure by Delta mode electrical measurement (≤0.1μΩ). Besides, transition temperature reduces to lower temperature by 4 K, after applying 1 tesla magnetic field perpendicular to the thin film. Few YouTube video links of temperature dependent electrical characterization of such a thin film is given next to the acknowledgement section.

## 1. Introduction

Science society has a century-long dream to develop materials that can show zero resistance (or extremely low resistance like few micro-ohms) at room or close to the room temperature because of their great potential for various applications including loss-less power distribution, energy storage, high power magnet development, and so on. This journey started from the discovery of superconductivity which was detected in the year of 1911 at the temperature 4.2 K.[1] From then, continuous efforts have been given to develop different materials that can show superconductivity at higher temperature. Most of those initial works have been done on different metals and alloys. During those developments in 1933, the Meissner effect shows that beside disappearing of resistivity, there is a transition of magnetic property of the material from paramagnetic to diamagnetic.[2] A number of theories such as London theory, Ginzburg-Landau theory, BCS theory have been proposed for the explanation of this phenomenon. Out of them the BCS theory proposed the formation of Cooper pair where two negative charge carriers (electron) are bounded via attractive force of electron-phonon-electron interaction which is considered one of the very unique and successful theory of solid state physics. Instead of that, after the discovery of high temperature superconductivity (High-$T_C$), BCS theory alone fails to explain this phenomenon.[3] Although, Cooper pair formation is still remains the fundamental issue because of its capability to form superconducting gap.[4, 5]

Again compared to bulk, thin film high-$T_C$ materials are more demanding for their versatile applications including superconducting circuit, supercomputing, SQUID etc.[6, 7] Earlier several theoretical predictions for achieving exceptionally high-$T_C$ thin film materials have been proposed, particularly on thin film based materials. Among them, the concept of surface superconductivity was proposed due to the existence of surface electrons of a thin film which are more localized to the crystal.[8, 9] Therefore, the interaction of the surface electrons of a thin film material is quite different from their bulk electrons.[10, 11] Again, interaction of those localized surface electrons can be modified by different 'impurity surface states' which can successfully develop a high $T_C$ superconductor.[12] During those studies, the possibility of achieving the superconducting phase of 1D polymer chain was discussed by Little et al.[13] According to him, overall center chain π-electron interaction of the polymer can be modulated by the side chain π-electron which can effectively overcome columbic repulsion. His proposal on attractive force of two electrons is known as the 'electron' mechanism which is different from conventional phonon mediated cooper pair formation concept. Later on, this electron mechanism was also adapted for high-$T_C$ thin film superconducting materials by several groups. In their model, a thin layer of metal film needs to be sandwiched between a dielectric film which is capable of enhancing $T_C$ by ~$10^2$ times over conventional phonon model.[14, 15] However, to achieve high-$T_C$ superconductor, the electron wave functions of metal and dielectric need to overlap which requires atomic distance separation between dielectric and metal.[16] Due to this limitation, this concept was not realized so far in a practical system. Beside this electron model, 'exciton model' has been proposed by Allender et al. where a thin layer of metal needs to be deposited on a semiconductor such a way that the Fermi energy of the metal exists in between valence and conduction band of the semiconductor. In such a situation, the metal electrons at the Fermi surface may tunnel into the semiconductor gap where they interact with virtual excitons, producing a net attractive interaction among the electrons.[17] Although, like the electron model, here also requires an intimate contact between metal and semiconductor which limits the realization of this concept for a practical material.[18]

Recently, Saha et al. has reported unconventional properties of engineered Au–Ag nanostructures. In that work, they are capable of growing Au nanostructures of size 10-60 nm embedded with uniformly distributed Ag nanoclusters of size ~1 nm, where separation of Ag and Au atom supposed to be in atomic scale because of their very similar lattice parameter. A thin film of such Au–Ag nanocrystals shows two distinct resistance levels below and above 245±1 K temperature with a variation of resistance ~$10^5$.[19] The lower level of resistance is ~ 2 μΩ which is the lower limit of their measurement set-up is believe to reach due to the superconducting transition of the film. A shifting of this switching towards room temperature was observed after successive cyclic temperature variation. Again, this transition temperature was shifted towards lower temperature after the application of magnetic field. Besides, a step like decrement in the voltage under constant current sourcing during transition was observed which is believed due to the formation of phase slip centre, an indication of superconducting transition. Additionally, it has been observed that these Au–Ag nanocrystals lose their conventional plamonic absorption.[20] In spite of discovering such unique results, the key limitation of their colloidal nanocrystal based materials is their very high air sensitivity which forces them to get those results with poor success rate.

In this work, an environmentally stable mesoscopic thin film of Au of a critical thickness has been deposited thermally on top of an $Ag^+$ implanted ion-conducting metal oxide (ICMO) thin film substrate. During this deposition a clean glass substrate and a ICMO coated substrate were kept as a references. This $Ag^+$ implanted substrate was prepared prior to the thermal evaporation through a solution processed technique. It has been observed that at room temperature, $Ag^+$ implanted Au thin film shows metallic conductivity (resistance of ~ 0.7 Ω) where as that reference film on glass substrate doesn't show any conductivity (resistance of ~ GΩ). More interestingly, during temperature dependent resistance (R-T) study, $Ag^+$ implanted Au thin film shows a sudden drop of resistance from 0.7 Ω to ≤0.1 μΩ at around 240 K. However, this transition temperature shifted close to the 0 °C after 3-4 successive cyclic variations of temperatures where it remained stable over the time. Detail description of this finding is given in the following section.

## 2. Experimental Section

### 2.1. Synthesis of materials.

As mention earlier, $Ag^+$ implanted substrate has been grown in a two-steps solution processed technique. Initially, a sol-gel approach is used to synthesize an ion-conducting metal Oxide (ICMO) ceramic thin film by precursor salt of metal oxide and lithium acetate (acquired from Alfa- Aesar, 99 % extra pure). In the beginning, certain concentration of metal oxide precursor and lithium acetate are prepared separately in 2-methoxy ethanol solvent under vigorous mixing at room temperature using magnetic stirrer until it forms clear solutions.

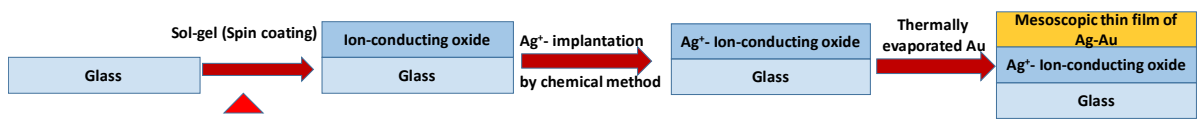

Figure.1 Schematic presentation of Ag-Au mesoscopic thin film deposition process

Following to this, these two solutions are mixed in a proper volume ratio, so that the final product can have the right stoichiometric ratio of that ICMO ceramic product. The mixture is then rapidly stirred for two hours.

## 2.2. Thin film Fabrication.

Initially a polycrystalline thin film of ICMO has been deposited on top of a glass substrate (15 mm x 15 mm) by a sol-gel technique followed by an annealing process. Thereafter, $Ag^+$ was implanted inside the ICMO thin film through a chemical process. Then a thin layer of Au with a critical thickness was deposited by thermal evaporation at room temperature on the top of this $Ag^+$ implanted ICMO thin film. During this Au deposition process, a clean glass substrate and a ICMO thin film coated substrate (without $Ag^+$) were kept as reference sample. This deposition method is a really very easy, low cost and a fast process. Using a spin coater, a muffle furnace and a thermal evaporator, it is possible to prepare even 500 pieces of such samples in a week with 100% reproducibility. Besides, this sample preparation, a thicker Au thin film has been deposited in the selective area of another $Ag^+$ implanted ICMO thin film by a shadow mask method. After that, a thin layer of Au (with critical thickness) is deposited in the entire area of substrate by removing the mask which allows us to obtain two different regions (thicker and thin) of Au thin film. Thereafter, resistance vs. temperature (R-T) measurement through four probe methods on two different samples have been investigated as described in the R-T measurement section.

## 3. Material Characterization.

Structural analysis of thin film samples is performed using X-ray diffraction (XRD) (Rigaku, Mini Flex600, DTEX Ultra) by using monochromatized Cu Kα radiation (λ= 1.54 Å) in the 2θ range of 10°–60° with a scan rate of 2°/min. Scanning electron microscope (SEM) study is done by using (NOVA NANOSEM 40, FEITM) to describe the surface morphology of the thin film. The elemental composition of the metallic elements is determined by an energy- dispersive X-ray spectrometer (EDX) attached to the HR-SEM with an accelerating voltage 10 kV. Atomic Force Microscopy (AFM) is carried out for the measurements of surface roughness of the samples in different concentration. All optical characterization (Transmittance and Reflectance) is studied by using EQE measurement unit (Enlitech QE-R). Photoluminescence spectra (PL) of the thin film has been studied by using HITACHI (F-4600) fluorescence spectrophotometer. For temperature dependent four probe electrical characterization, sample was placed in contact with 4-spring loaded equidistance probes (Mill Max 858-22-004-30-001101 4POS, 4MM, SMD) attached with cryogenic measurement set-up (Physical Quantity Measurement System (PQMS), XPLORE). These probes are coated with gold to minimize contact resistance with the sample. Resistance of the samples are measured with three different electronics instruments; digital multimeter (Keithley DMM6500 6.5 Digit Multimeter), dual soucemeter (Keysight B2912A) and combined current sourcemeter and a nanovoltmeter (Tektronix 6221+2182A, Delta mode measurement) in 4-probe configurations where as voltage vs. current (V vs. I) and time dependent voltage (V vs. t) under constant current sourcing has been performed by using Delta mode measurement.

## 4. Results and Discussion

### 4.1 UV-VIS absorption and Photoluminescence (PL) spectra of thin films

UV-VIS spectra of Ag-Au mesoscopic thin film has been demonstrated in Fig. 2a (red) which shows a strong absorption at 740 nm due to the plasmonic effect of Ag-Au nanocrystal. Whereas, the absorption peak of reference Au film on glass is ~864 nm and Ag thin film (same thickness as of Au) is ~ 510 nm. This comparative data implies that the plasmonic absorption of Ag-Au mesoscopic thin film is due to the combined effect of Ag and Au nanoparticles.

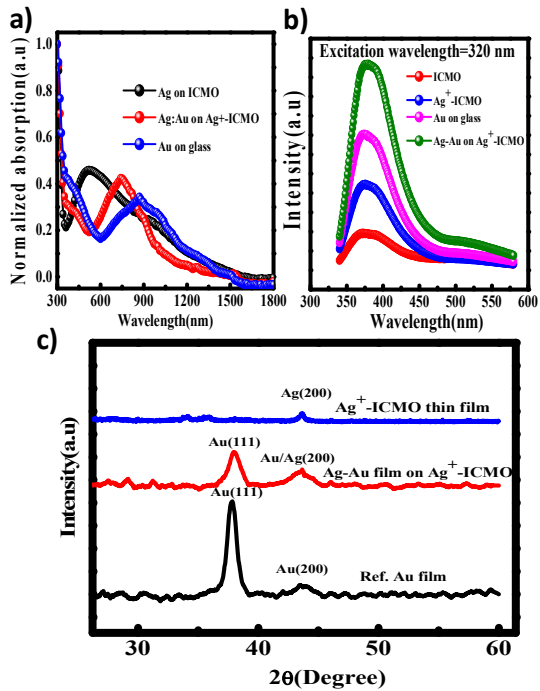

Figure 2. a) UV-VIS absorption spactra of Ag on ICMO (black), Ag-Au mesoscopic thin film (red) and reference Au thin film (blue), b) photoluminescence (PL) spectra of ICMO (red), Ag$^+$ implanted ICMO (blue), reference Au film on glass (pink), and Ag-Au mesoscopic thin film (green), c) XRD patterns of Ag$^+$ implanted ICMO (blue), Ag-Au mesoscopic thin film (red) and reference Au thin film (black).

Photoluminescence (PL) spectra of respective films have also been investigated with excitation wavelengths of 320 nm. Original ICMO thin film has been used as reference for this study. It has been observed that the PL signal of ICMO thin film is very weak (Fig 2b, red). However, after Ag$^+$ implantation, PL signals enhance significantly (Fig 2b, blue), although this PL intensity is weaker than reference Au film on glass (Fig 2c, pink). Most likely, this enhancement of PL after Ag$^+$ implantation originated due to some Ag cluster formation during this Ag$^+$ implantation process. After Au deposition on Ag$^+$ implanted ICMO, PL intensity of the film enhanced more than double (Fig 2b, green). Although the nature of PL spectra in all of these films are quite similar with a peak position at ~375 nm.

### 4.2 X-ray diffraction (XRD)
Figure 2c) shows the XRD spectra of different stage of Ag-Au thin film fabrication. Thin film XRD samples are made through spin coating and a subsequent annealing process on a p$^+$-Si substrates. Due to the amorphous nature of glass, we choose p$^+$-Si substrate for XRD study to reduce the background signal. Using a Rigaku X-ray diffractometer with Cu-K$\alpha$ radiation ($\lambda$= 1.54 Å) in the 2$\theta$ range of 10°–60° at a scan rate of 2°/min, the crystallinity and phase of these thin films materials are studied. The X-ray diffraction (XRD) spectra of the thin films are acquired at room temperature. In the thin-film XRD pattern of Ag$^+$-ICMO shows a tiny peak formation at 2$\theta$ ~ 43.7° which is corresponding to the (200) plane of Ag (JCPDS file no 897322). This data indicates that during Ag$^+$ implantation process, few Ag cluster is also formed. The XRD pattern of Ag-Au film on Ag$^+$-ICMO is shown in Fig. 2c (red) which shows two intense peak at 2$\theta$ ~38.0° and 43.7° respectively which are correspond to the planes of (111) and (200) of Au respectively (JCPDS file no. 04-0784). These peak positions are common for Ag as well because of their very similar lattice parameters. From this picture its very clear that the width of both peaks are quite wide, indicates the small size of the metal nanoparticle. In contrast, XRD data of reference Au film which is deposited on bare Si-substrate, has same peak position but with narrow width.

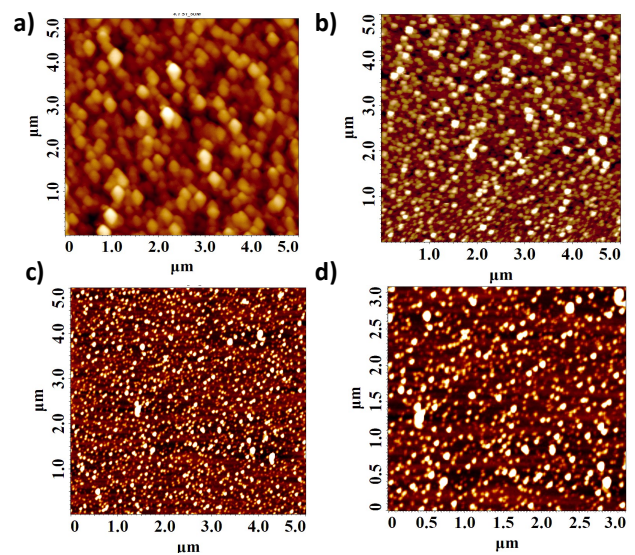

Fig.3 The AFM images of a) ICMO thin film b) Ag$^+$ implanted ICMO thin film c)-d) Ag-Au mesoscopic thin film

### 4.3 Atomic Force Microscopy (AFM) of Ag-Au mesoscopic thin film
Atomic Force Microscopy (AFM) study has performed on the ICMO, Ag$^+$ implanted ICMO and Ag-Au mesoscopic thin films which is shown in Fig. 3. The AFM picture of ICMO thin film indicates formation of contentious film of polycrystalline grain of average grain size ~ 30 nm (Fig. 3a) with rms roughness of ~3.5 nm. However, after Ag$^+$ implantation on that ICMO thin film through a chemical process, surface morphology of this film totally changes and form much smaller grains as shown in Fig. 3b) with average grain size of this film is ~22 nm whereas, rms roughness of this film is ~ 6.8 nm. After thin layer of Au deposition on top of this Ag$^+$ implantation ICMO thin film, a percolated continuous metal nanoparticle formation occurs which is seen as bright circular spot in the image. The rms roughness of this film is ~ 4.1 nm which is lower than Ag+ implantation ICMO thin film. Due to the similar lattice parameter of Ag and Au, possibly Ag-cluster/Ag+ of ICMO thin film has been works as nucleation site of Au nanoparticles, selectively on those sites. After certain amount of Au deposition, these metal nanoparticle becomes percolated that makes the film conducting. Possibly, during the growth of Au nanoparticle, Ag+ converted to tinny Ag cluster and reside inside the bigger Au nanoparticle to form a Ag-Au nanocrystal.

### 4.4 Scanning electron microscopy (SEM) of Ag-Au mesoscopic thin film

Scanning electron microscopy (SEM) is being used to study surface morphology and elemental composoition of the Ag-Au mesoscopic thin film which is shown in Fig. 4. SEM picture of this film indicates (Fig. 4a, inset) that the film has some crackes of width ~20-30 nm, but its percolation is maintained which makes it a conducting film. The energy dispersive X-ray spectroscopy (EDS) elemental maps of Au (Fig. 4c), Ag (Fig. 4d) of the film indicates that both Ag and Au has been deposited everywhere of the film. However, ratio of Ag and Au is ~ 1:20.

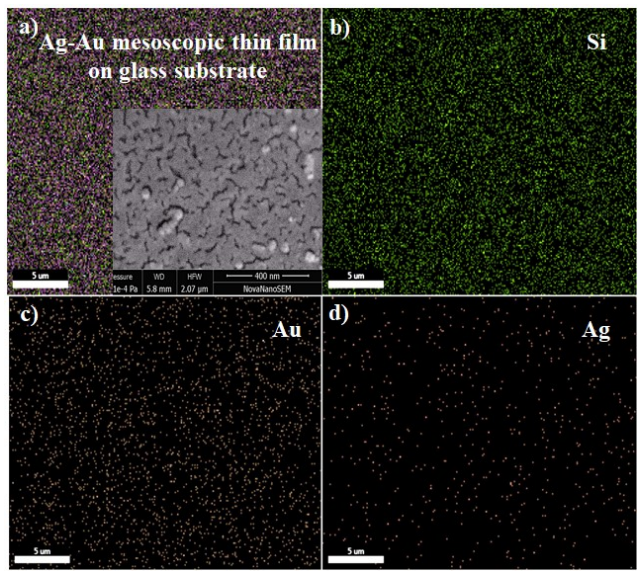

Fig 4. Elemental mapping analysis from SEM photographs (a) combinaed picture for all elements (SEM picture in the inset), b) for Si, c) for Au and d) for Ag

### 4.5 Possible growth of Ag claster embaded Au nanoparticle

Our XRD data of Ag+ implanted ICMO thin film shows a weak signal of Ag, indicates silver cluster formation after Ag+ implantation inside ICMO. Agin, a enhanced photoluminescence has been observed of this Ag+ implanted ICMO thin film at ~400 nm which is also the indication of Ag cluster formation. Moreover, this PL signal becomes more intense when Au is themally evaporate on it. Since, Ag and Au crystal have very similar lattice parameter, therefore during Au deposition, pre-existing Ag-cluster/Ag+ on the substrate works as nucleation site of Au nanoparticle. However, due to the room temperature deposition of Au thin film, chance of Au-Ag alloys formation is very less. Relatively, Ag cluster embaded Au nanoparticle formation is more possible pheneomena. To ensure that, more detail microstructural analysis is required in future. Although, Au film on a clean glass substrate and original ICMO thin film coated substrate remain non-conducting under same batch of Au deposition which is due to the lack of percolation path of Au. The difference of conductivity of Au film over Ag+ implanted ICMO film with other substrates implied the crucial role of Ag-cluster/Ag+ during Au deposition. This phenomenon is also realized from the difference of the surface morphology of these Au thin films which has been checked from AFM study (AFM picture is not added).

### 4.6 Electrical characterization of Ag-Au mesoscopic thin film

In the temperature dependent four probe measurement, we used gold coated equally separated align four spring loaded pogo pin connectors (Fig. 5a, MILL MAX 858-22-004-30-001101) for making contact with the film. This pogo pin connector was attached through two adjustable screws with the sample holder of a cryocooler as shown in Fig. 5a). Electrical resistance was measured by in three different ways by using three sets of independent instruments; Keysight dual source meter (4-wire mode), Delta Mode Measurements (using Tektronix nanovoltmeter in combination of current sourcemeter) and Tektronix digital multi-meter (4-wire mode).

*4.6a Resistance Vs. Temperature (R-T) data*

Room temperature resistance of such Ag-Au mesoscopic thin film that measured by this 4-probe arrangement is ~ 0.7 Ω (Fig. 5b). As temperature reduces, film resistance also reduces slowly. However, around 240 K, the sample reduces its resistance suddenly by 6-7 orders of magnitude within 0.1 K variation of temperature and reaches to ≤0.1 μΩ (Fig. 5c) which is the noise floor of these instruments. We hardly find more than one data in between these two levels of resistance. If we reduce temperature further, it continuously shows its low resistance level. During increasing temperature, low-to-high resistance transition occurs at ~ 245 K. Interestingly, if we do this temperature cycling, in the second cycle it shows high-to-low transition at ~250 K

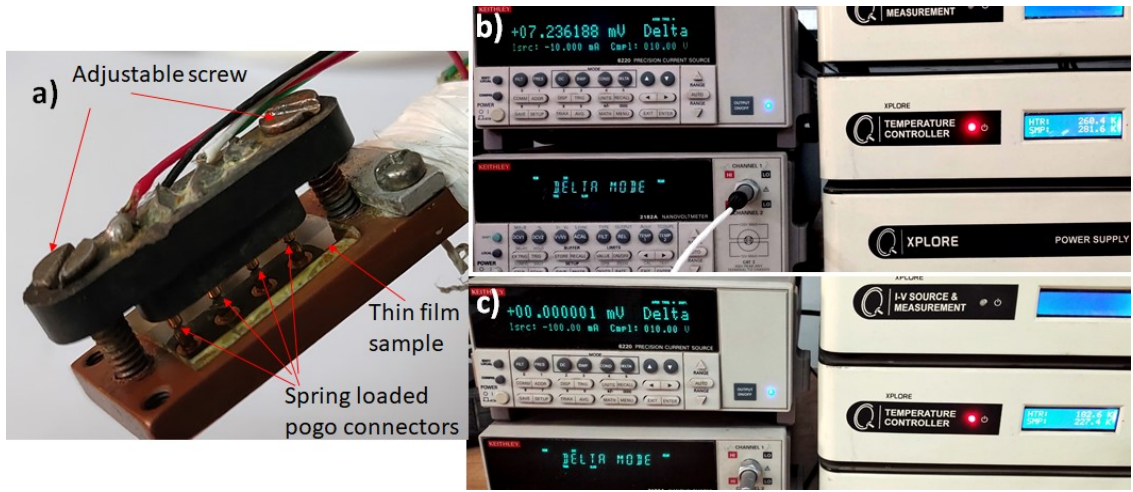

Fig. 5 a) Picture of spring loaded pogo connector sample holder for 4-wire measurement, b) resistance of the sample at 281.6 K, c) resistance of the sample at 227.4 K and low-to-high at ∼ 255 K. After three successive cycles, $T_C$ becomes close to 0 °C and remains stable for months (Fig. 6a). In our last 30 samples, we found almost similar data and didn't find any failure in any single sample. At this stage, the reason behind this shifting of $T_C$ with initial temperature cycling is not clear to us and needs further investigation in future. Again, we store our samples in open atmospheric condition, which doesn't change its behaviour after months. During this study, we also measure the R-T of the two region of Au film (thick and thin) of the 2nd sample which is also deposited on a $Ag^+$ implanted ICMO substrate (Fig. 6b, inset). Interestingly, thinner Au area (with critical thickness) shows resistance drop with reducing temperature as like the initial sample. However, a thicker portion of Au doesn't show any transition up to 150 K, indicating the requirement of critical thickness of Au for this transition (Fig. 5b). This result is really very exciting for the fabrication purpose of a superconducting circuit. Because, just by selective thick and thin (with critical thickness) Au deposition on $Ag^+$ implanted ICMO substrate with several μm separation, it is possible to fabricate Josephson junction very easily. This μm width 'thick Au' can be deposited by lithographic process on $Ag^+$ implanted ICMO substrate prior to the critical thickness Au deposition, to fabricate verities of superconducting circuit which can work close to the room temperature.

***4.6b Voltage drop after transition at different current level and polarities (V vs I)***
After transition, (in low resistance phase) we did systematic study of voltage drop across the voltage-probe at different current intensity, ranging from -100mA to +100 mA by using Delta mode measurement (nanovoltmeter in combination with a current sourcemeter) set up (7:00 min to 17:10 min of YouTube video). During this study, we didn't find any variation of voltage drop within the entire scale including two different polarities of current, indicating resistance of this sample is too low to detect under such circumstances. Again, we measured voltage drop by turning on and off the current sauce and didn't find any variation of voltage fluctuation. In the entire studies, we found only a few nV voltage fluctuation regardless the current level. The fluctuation of nanovoltmeter data during sweeping current from -100 mA to +100 mA is shown in the V vs. I characteristics (Fig. 5c), indicating that voltage drop is always <±10nV in the entire range. Again, this V-I characteristics is very much random in different independent measurements, although the value of voltage drop always remains within ±10nV.

***4.6c Voltage drop after over the time at different current polarities (V vs t)***

After superconducting transition of Ag-Au mesoscopic thin film, voltage drop across the voltage probe has been studied over the time (V vs. t). During this measurement, +100 mA current was fixed in the current probe for one set of measurement (2:17 min to 4:18 min of YouTube video) and continue that for 2 minutes. In a subsequent measurement, -100 mA current was set in the current probe and continue for two minutes as well. The variation of voltage across voltage probe over the time (V Vs. t) is shown in Fig. 6d (4:33 min to 6:36 min of YouTube video). This data indicates that voltage drop is entirely fluctuating within ±10 nV regardless the current polarity of the current probe, indicating that the sample resistance remain below the limit of this measurement range (100 mA/10nV=0.1 μΩ).

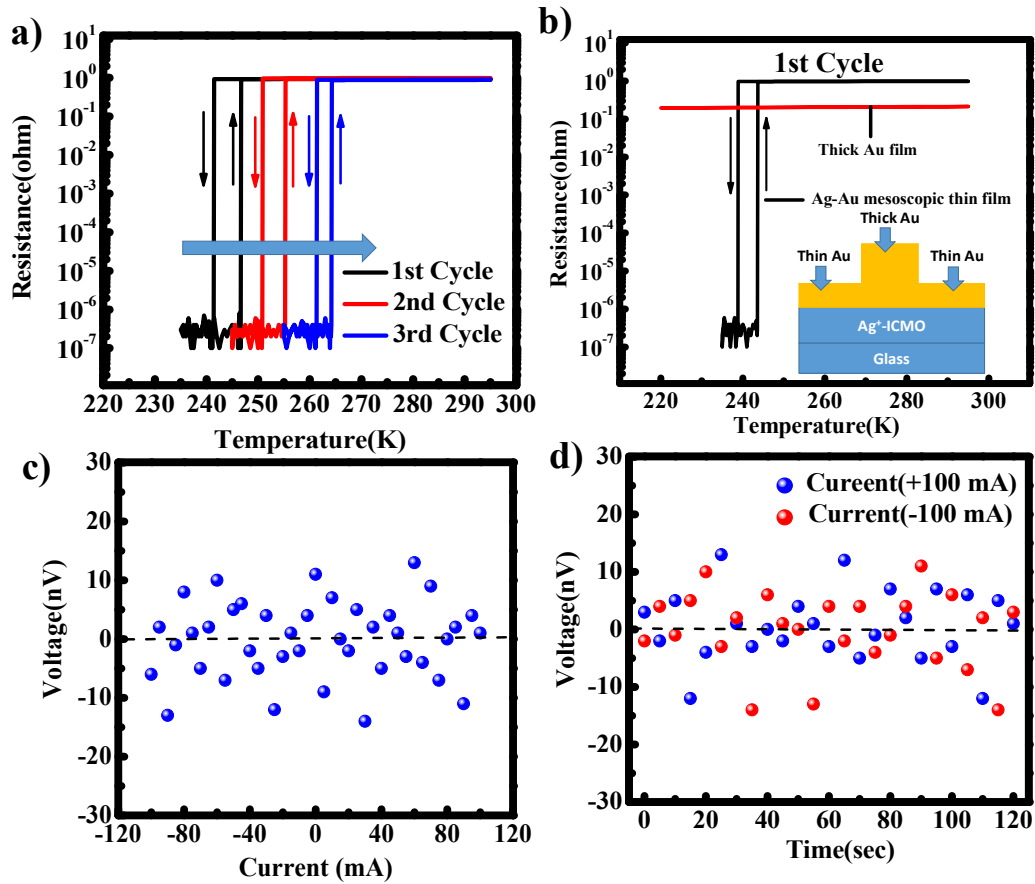

Fig. 6 a) The R-T data with different temperature cycles of Ag-Au mesoscopic thin film, b) comparative R-T data of thick and thin region of Au film of sample 2, inset shows the schematic diagram of selective thick and thin layer of Au c) V vs I data after low resistance transition (~256 K), d) current vs. time data (I vs. t) under +100 mA and -100 mA current after low resistance transition (~256 K)

*4.6d Magnetic field dependent transition ($T_C$)*

The R-T measurement of Ag-Au mesoscopic thin film has been studied under magnetic fields as well. During this study, a magnetic field was applied perpendicular to the substrate which has been varied from 0 to 1 Tesla to realize the effect of magnetic field on transition temperature. This measurement has been performed with a sample on which three temperature cycle R-T measurement has been performed earlier. The R-T data under magnetic field has been shown in Fig. 6d) which indicates that under 1 T magnetic field, the transition temperature shifted from 263 K to 259 K without changing the nature of transition, implying that the vertical magnetic field has some effect on transition temperature, but can't destroy superconducting behavior so easily. Although, due to the limitation of our facility we are unable to check this effect at higher magnetic fields.

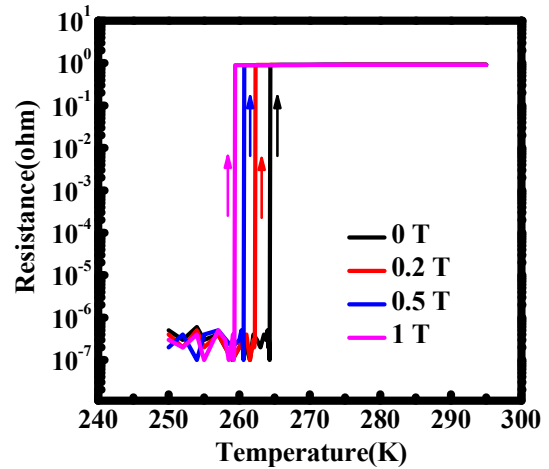

Fig. 7 Magnetic field dependent R-T data of Ag-Au mesoscopic thin film

## 5. Conclusion

A stable Ag-Au mesoscopic thin film has been deposited in a very low cost, easy and fast method that shows a close to the room temperature thin film superconductivity. For this deposition, a thin layer of Au has been deposited on top of a $Ag^+$ implanted ICMO thin film. Superconducting transition temperature of such film initially observed at ∼240 K which shifted close to $0°$ C after 3 successive temperature cycles and remained stable over the time. At low resistance state, current-voltage data within ±100 mA is randomly fluctuating within ±10 nV range, indicating sample resistance becomes ≤0.1μΩ. This transition temperature also shifted towards a more negative direction by applying a magnetic field perpendicular to the film. Although this shifting is only 4 K after applying 1T magnetic field and doesn't change the nature of transition. Again, a thicker Au film on the same $Ag^+$ implanted substrate doesn't show any such transition which has been tested up to 150 K, indicating, a superconducting circuit can be easily fabricated on this $Ag^+$ implanted substrate by selective thick and thin Au deposition only. This selective thick Au deposition or patterning can be done by lithography process prior to the thin Au deposition to fabricate verities of the superconducting circuit which can work close to the room temperature.

## Acknowledgements

Bhola Nath Pal thanks SERB, India (CRG/2019/001826) and DST, India (DST/INT/SWD/VR/P-12/2019) for financial support. Authors are also grateful to Central Instrument Facility Centre, IIT (BHU), for providing instrument support for XRD, AFM and SEM. Sobhan Hazra is thankful to SERB for providing Ph.D. fellowship.

## YouTube Video link of electrical characterization

**1. 3rd cycle, complete cycle of R-T, V-I, V-t data**

https://youtu.be/nlEUCWWTZuM

- Superconducting phase transition from Higher to Lower Resistances @ 1:29 min of the video
- Voltage vs Time at constant current (+100 mA) from 2:17 min to 4:18 min of the video
- Voltage vs Time at constant current (-100 mA) from 4:33 min to 6:39 min of the video
- Voltage vs Current with different current sweep (+100 mA to -100 mA) with interval of 10 mA from 7:00 min to 17:10 min of the video
- Superconducting phase transition from Lower to Higher Resistances @28.04 min of the video

**2. 1st Temperature cycle of R-T, from high to low resistance phase transition @ 3:10 min of the video**

https://youtu.be/6Ya1EqVks1I

**3. 1st temperature cycle of R-T, low to high resistance phase transition @4:40 min of the video**

https://youtu.be/JlMyBz7MT9k

**4. 2nd cycle phase transition, height to low resistance @2:15 min of the video**

https://youtu.be/Re741xYi3nM

**5. 2nd temperature Cycle, low to high resistance phase transition@4:45 min of the video.**

https://youtu.be/hpc06nRy-eQ

## Authors contribution

**Bhola Nath Pal:** Conceptualization, draft writing/editing and supervising the project

**Sobhan Hazra:** Most of the experimental work, draft editing

**Sandip Chatterjee:** Support magnetic field dependent R-T measurement and draft editing